\documentclass[
superscriptaddress,
nofootinbib,
twocolumn,
 amsmath,amssymb,
 aps,
pra,
notitlepage,
longbibliography
]{revtex4-1}

\usepackage{dcolumn}
\usepackage{bm}
\usepackage{epsfig}
\usepackage{graphicx}
\usepackage{latexsym}
\usepackage{amsfonts}
\usepackage{float} 
\usepackage{setspace}
\usepackage{graphicx}
\usepackage{amsmath}
\usepackage{mathrsfs}
\usepackage{verbatim}
\usepackage{color}
\usepackage{SIunits}
\usepackage{hyperref}
\usepackage{physics}

\newcommand{\be}{\begin{equation}}
\newcommand{\ee}{\end{equation}}
\newcommand{\ba}{\begin{array}}
\newcommand{\ea}{\end{array}}
\newcommand{\bqa}{\begin{eqnarray}}
\newcommand{\eqa}{\end{eqnarray}}

\begin{document}
\title{InGaP $\chi^{(2)}$ integrated photonics platform for broadband, ultra-efficient nonlinear conversion and entangled photon generation }

\author{Joshua Akin} 
\thanks{These authors contributed equally to this work.}
\affiliation{Holonyak Micro and Nanotechnology Laboratory and Department of Electrical and Computer Engineering, University of Illinois at Urbana-Champaign, Urbana, IL 61801 USA}
\affiliation{Illinois Quantum Information Science and Technology Center, University of Illinois at Urbana-Champaign, Urbana, IL 61801 USA}
\author{Yunlei Zhao} 
\thanks{These authors contributed equally to this work.}
\affiliation{Holonyak Micro and Nanotechnology Laboratory and Department of Electrical and Computer Engineering, University of Illinois at Urbana-Champaign, Urbana, IL 61801 USA}
\affiliation{Illinois Quantum Information Science and Technology Center, University of Illinois at Urbana-Champaign, Urbana, IL 61801 USA}
\author{Yuvraj Misra} 
\affiliation{Holonyak Micro and Nanotechnology Laboratory and Department of Electrical and Computer Engineering, University of Illinois at Urbana-Champaign, Urbana, IL 61801 USA}
\affiliation{Illinois Quantum Information Science and Technology Center, University of Illinois at Urbana-Champaign, Urbana, IL 61801 USA}
\author{A. K. M. Naziul Haque} 
\affiliation{Holonyak Micro and Nanotechnology Laboratory and Department of Electrical and Computer Engineering, University of Illinois at Urbana-Champaign, Urbana, IL 61801 USA}
\affiliation{Illinois Quantum Information Science and Technology Center, University of Illinois at Urbana-Champaign, Urbana, IL 61801 USA}
\author{Kejie Fang} 
\email{kfang3@illinois.edu}
\affiliation{Holonyak Micro and Nanotechnology Laboratory and Department of Electrical and Computer Engineering, University of Illinois at Urbana-Champaign, Urbana, IL 61801 USA}
\affiliation{Illinois Quantum Information Science and Technology Center, University of Illinois at Urbana-Champaign, Urbana, IL 61801 USA}

\begin{abstract} 

Nonlinear optics plays an important role in many areas of science and technology. The advance of nonlinear optics is empowered by the discovery and utilization of materials with growing optical nonlinearity. Here we demonstrate an indium gallium phosphide (InGaP) integrated photonics platform for broadband, ultra-efficient second-order nonlinear optics. The InGaP nanophotonic waveguide enables second-harmonic generation with a normalized efficiency of $128,000\%$/W/cm$^2$ at 1.55 $\mu$m pump wavelength, nearly two orders of magnitude higher than the state of the art in the telecommunication C band. Further, we realize an ultra-bright, broadband time-energy entangled photon source with a pair generation rate of 97 GHz/mW and a bandwidth of 115 nm centered at the telecommunication C band. The InGaP entangled photon source shows high coincidence-to-accidental counts ratio CAR $>10^4$ and two-photon interference visibility $>98\%$. The InGaP second-order nonlinear photonics platform will have wide-ranging implications for non-classical light generation, optical signal processing, and quantum networking.

\end{abstract}

\maketitle

\noindent\textbf{Introduction}\\
The development of nonlinear optics is empowered by the invention of nonlinear materials, from bulk nonlinear crystals and silica fibers to more recent wafer-scale thin-film materials. Over the past decades, the application of materials with increasing nonlinearities, combined with the advance of light-confining nanophotonic structures, has resulted in a remarkable enhancement in nonlinear optical efficiencies. For example, the second-harmonic generation has advanced from the initial demonstration using a quartz crystal with a $10^{-9}\%/$W efficiency \cite{franken1961generation} to the record of $10^{5}-10^{6}\%/$W achieved in thin-film nanophotonic resonators nowadays \cite{lu2020toward,zhao2022ingap}.

\begin{figure*}[!htb]
\begin{center}
\includegraphics[width=2\columnwidth]{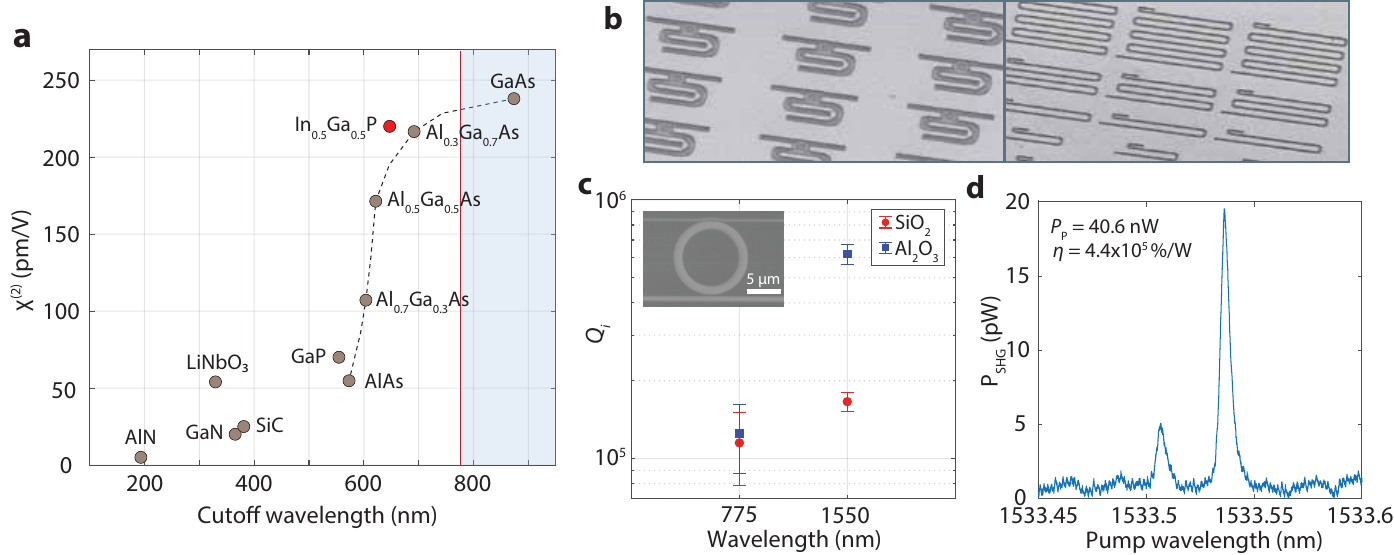}
\caption{ \textbf{a}. Second-order susceptibility and cutoff wavelength of several nonlinear optical thin-film materials. The red line indicates the 775 nm wavelength. Shaded region indicates optical absorption at wavelengths longer than 775 nm. AlN: \cite{pernice2012second}, LiNbO$_3$: \cite{wang2017second}, GaN: \cite{sanford2005measurement}, SiC: \cite{sato2009accurate}, GaP: \cite{shoji1997absolute}, AlGaAs: \cite{ohashi1993determination}, GaAs: \cite{shoji1997absolute}, InGaP: \cite{ueno1997second}. \textbf{b}. SEM images of InGaP photonic integrated circuits. \textbf{c}. Intrinsic quality factor of the 1550 nm TE$_{00}$ and 775 nm TM$_{00}$ resonances of $R=5\ \mu$m InGaP microring resonators (inset) with SiO$_2$ and Al$_2$O$_3$ claddings. 
\textbf{d}. Second-harmonic generation in a $R=5\ \mu$m phase-matched InGaP microring resonator with Al$_2$O$_3$ cladding. On-chip pump power is 40.6 nW and the peak SHG efficiency is 440,000 \%/W.}
\label{fig:platform}
\end{center}
\end{figure*}

Second-order ($\chi^{(2)}$) optical nonlinearity, as the dominant optical nonlinearity, enables a variety of nonlinear optical processes with high efficiencies and low noises, including generation of entangled photons \cite{zhao2020high} and squeezed light \cite{vahlbruch2016detection}, parametric optical amplification \cite{ledezma2022intense}, and coherent wavelength conversion \cite{guo2016chip}.
Fig. \ref{fig:platform}a displays the second-order susceptibility and cutoff wavelength of a selection of $\chi^{(2)}$ materials that are available in thin-film platforms.
Among them, III-V semiconductors, including GaAs and Al$_x$Ga$_{1-x}$As, are notable for the very high second-order susceptibility, leading to a long history of study for nonlinear optics \cite{vyas2022group}. The versatile III-V photonics platform enables heteroeptaxial integration of pump lasers and photodetectors, which is unique compared to other platforms. 
However, one drawback of these III-V semiconductors is the optical losses at short wavelengths. For example, GaAs has a narrow bandgap corresponding to a cutoff wavelength of 872 nm. While Al$_x$Ga$_{1-x}$As exhibits a wider bandgap, its second-order susceptibility decreases drastically with the increasing aluminum composition \cite{ohashi1993determination}. 
Moreover, arsenic III-V materials suffer from strong optical absorption at wavelengths less than 800 nm, due to the antibonding As-As surface state that is below the bandgap 
\cite{michael2007wavelength,lin2012passivation, placke2024telecom}.  These facts have limited the use of arsenic III-V materials for efficient second-order nonlinear optics in the important telecommunication C band (1530-1565 nm), where long-haul optical communications conducts, due to the absorption of the corresponding second harmonics.

Indium gallium phosphide (In$_{0.5}$Ga$_{0.5}$P, hereafter referred to as InGaP) is another III-V semiconductor material that is lattice-matched with GaAs and thus can be epitaxially grown on the GaAs substrate at the wafer scale. Because of its high electron mobility, direct bandgap, and thermal stability, InGaP has been used for making heterojunction bipolar transistors \cite{pan1998high}, solar cells \cite{takamoto2005ingap}, photodetectors \cite{jiang2004high}, and LEDs \cite{svensson2008monolithic}.  More recently, thin-film InGaP has been explored for third-order nonlinear optics using its substantial Kerr nonlinearity \cite{eckhouse2010highly, colman2010temporal}, resulting in demonstrations of frequency combs \cite{dave2015dispersive}, optical parametric oscillators \cite{marty2021photonic}, and entangled photon pairs via four-wave mixing \cite{chopin2023ultra}.

Besides its notable Kerr nonlinearity, InGaP is particularly appealing for second-order nonlinear optics because of the combination of a substantial second-order susceptibility ($\chi^{(2)}_{xyz} \approx$ 220 pm/V \cite{ueno1997second}) and a sizable bandgap of 1.92 eV (cutoff wavelength 645 nm). For InGaP, the antibonding anion state lies well above the bandgap \cite{lin2012passivation}, in contrast to Al$_x$Ga$_{1-x}$As, which avoids light absorption before the bandgap. InGaP also has a large refractive index ($>3$), which facilitates nanophotonic structures with strong light confinement. These properties suggest the potential of InGaP for realizing highly-efficient second-order nonlinear optics, in particular, in the crucial telecommunication C band. 
Recently, several research groups have embarked on studying second-order nonlinear optics using thin-film InGaP \cite{poulvellarie2021efficient,peralta2022low,zhao2022ingap}. Despite demonstrating a record nonlinearity-to-loss ratio in InGaP microring resonators \cite{zhao2022ingap}, most demonstrations thus far are still limited by considerable optical losses and imperfect phase matching condition. Moreover, InGaP microcavities demonstrated in Ref. \cite{zhao2022ingap} are unsuitable for applications that demand broadband operation and high optical powers. As a result, realizing a broadband, low-loss, ultra-efficient second-order nonlinear photonics platform based on InGaP remains elusive.

Here, through the optimization of optical losses and phase-matching condition for InGaP nanophotonic waveguides across an octave wavelength span, we demonstrate a broadband, ultra-efficient InGaP second-order nonlinear photonics platform in the telecommunication band. The demonstrated second-harmonic generation with a normalized efficiency of 128,000$\%$/W/cm$^2$ in the telecommunication C band is nearly two orders of magnitude more efficient than the state of the art \cite{wang2018ultrahigh,chen2024adapted}. Using the InGaP nanophotonic waveguide, we demonstrate an ultra-bright time-energy entangled photon source with a pair generation rate of 97 GHz/mW and a bandwidth of 115 nm (14.4 THz) centered at the telecommunication C band.
The broadband, ultra-efficient InGaP nanophotonics platform will enable a wide range of nonlinear optical processes and applications with unprecedented performances.

\begin{figure*}[!htb]
\begin{center}
\includegraphics[width=2\columnwidth]{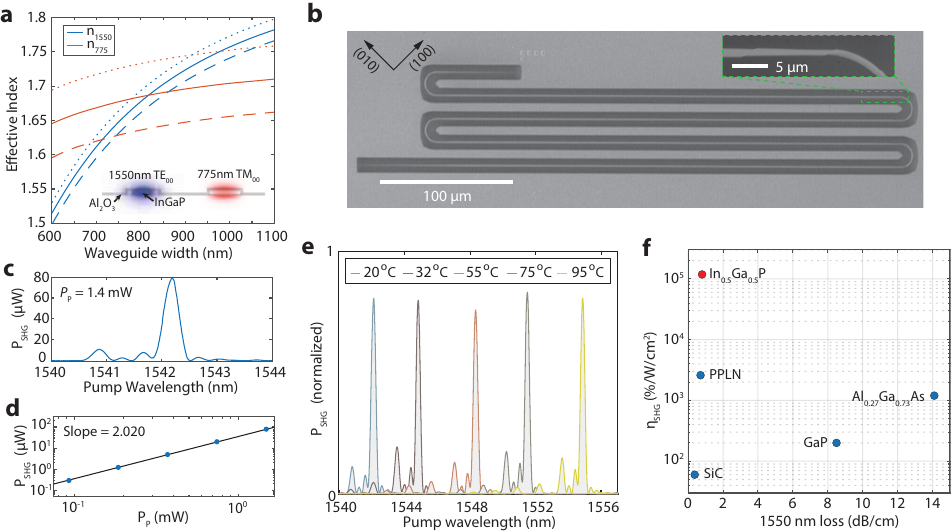}
\caption{\textbf{a}. Simulated effective mode index of 1550 nm TE$_{00}$ and 775 nm TM$_{00}$ modes of 108 nm (dashed), 110 nm (solid), and 112 nm (dotted) thick InGaP waveguides with 35 nm Al$_2$O$_3$ cladding. Inset is a cross section of the waveguide along with the field distribution of the two modes. \textbf{b}. SEM images of a fabricated InGaP waveguide device. \textbf{c}. Measured SHG spectrum for a 1.6 mm long waveguide device with a pump power of 1.56 mW. \textbf{d}. Quadratic relation between the peak SHG power and the pump power. \textbf{e}. Temperature tuning of the phase-matching condition. \textbf{f}. Normalized SHG efficiency and 1550 nm wavelength band loss of the  telecommunication C band waveguide SHG of several thin-film second-order nonlinear photonics platforms. InGaP: this work, PPLN: \cite{wang2018ultrahigh}, Al$_{0.27}$Ga$_{0.73}$As: \cite{may2019second}, GaP: \cite{pantzas2022continuous}, SiC: \cite{zheng2022efficient}.  }
\label{fig:SHG}
\end{center}
\end{figure*}

\noindent\textbf{Results}\\
In this study, 110 nm thick InGaP is epitaxially grown on the GaAs substrate (0 degree off-cut toward [110]) using metal-organic chemical vapor deposition (T=545 C, V/III=280, with precursors including trimethylindium, trimethylgallium, and PH$_3$). 
The root-mean-square surface roughness of the InGaP thin film is measured to be about 0.3 nm, which is close to the native surface roughness of the GaAs substrate. To fabricate InGaP photonic integrated circuits, bonding of InGaP thin film to low-index substrates have been demonstrated before \cite{dave2015nonlinear,peralta2022low}. Here we adopted a transfer-free approach to fabricate InGaP photonic integrated circuits with low-index oxide top claddings (Methods) \cite{zhao2022ingap}. Fig. \ref{fig:platform}b shows the scanning electron microscopy (SEM) images of fabricated InGaP photonic integrated circuits, including microring resonators and waveguides. 

We studied the optical loss of InGaP nanophotonic devices in both 1550 nm and 775 nm wavelength bands with two different oxide claddings, SiO$_2$ and Al$_2$O$_3$, deposited via atomic layer deposition. The optical loss is characterized using the intrinsic quality factor ($Q_i\equiv\frac{\omega}{\kappa_i}$) of InGaP microring resonators. The microring resonator couples with both 1550 nm and 775 nm wavelength-band waveguides, which enable measurement of the transmission spectrum of the device \cite{zhao2022ingap}. $Q_i$ is then inferred from the fitting of the resonance spectrum. Fig. \ref{fig:platform}c shows the measured $Q_i$ of the 1550 nm band fundamental transverse-electric (TE$_{00}$) resonance and the 775 nm band fundamental transverse-magnetic (TM$_{00}$) resonance of microring resonators with 5 $\mu$m radius and 1 $\mu$m width. The average value of $Q_{i, 1550}$ for microring resonators with Al$_2$O$_3$ cladding is about $6\times 10^5$, over three times higher than that with SiO$_2$ cladding. $Q_{i, 775}$ is $1-2\times 10^5$ and shows slight improvement with Al$_2$O$_3$ cladding. The increase of the quality factor could be attributed to the surface passivation induced by Al$_2$O$_3$ \cite{guha2017surface}. We also made microrings with different sizes and found $Q_{i, 1550}$ is peaked around $8\times 10^5$ for larger rings, which is limited by the absorption loss (Supplementary Information (SI)). Leveraging the optimized optical loss, we measured second-harmonic generation (SHG) in a 5-$\mu$m-radius ring with phase-matched 1550 nm band TE$_{00}$ and 775 nm band TM$_{00}$ resonances and realized a resonant nonlinear conversion efficiency $\eta\equiv P_{\mathrm{SHG}}/P_p^2=440,000\%$/W (Fig. \ref{fig:platform}d). This represents a 6-fold enhancement over the previously reported value using the same InGaP microring resonator but with SiO$_2$ cladding \cite{zhao2022ingap}.  

\begin{figure*}[!htb]
\begin{center}
\includegraphics[width=2\columnwidth]{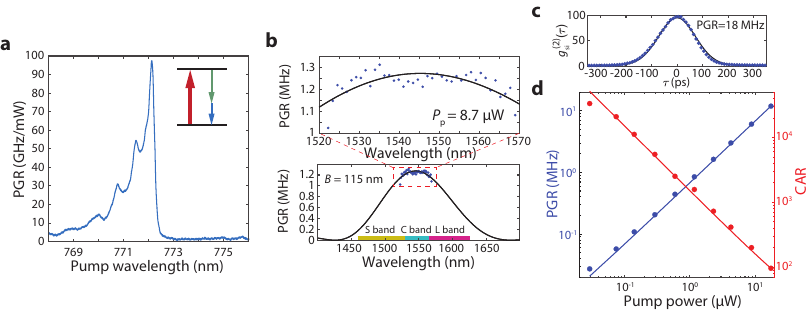}
\caption{\textbf{a}. Dependence of the total SPDC pair generation rate (PGR) efficiency on the pump wavelength. \textbf{b}. SPDC spectrum sampled by a DWDM with $P_p=8.7\ \mu$W. The solid line is a sinc$^2$ function fitting, yielding a bandwidth of 115 nm (14.4 THz).  \textbf{c}. Measured $g_{\mathrm{si}}^{(2)}(\tau)$ for $\mathrm{PGR}=18$ MHz in one DWDM channel. Solid line is Gaussian fitting. \textbf{d}. Pair generation rate in one DWDM channel and coincidence-to-accidental counts ratio (CAR) for various pump powers. Solid lines are linear fitting. }
\label{fig:SPDC}
\end{center}
\end{figure*}

In contrast to cavities, waveguides can be operated in the broadband regime and circumvent the light extraction issue associated with cavities. Similar to the microring resonator, the InGaP waveguide is designed for phase matching between the 1550 nm band TE$_{00}$ mode $a$ and the 775 nm band TM$_{00}$ mode $b$, which satisfy $2\omega_{a}=\omega_{b}$ and $2k_{a}=k_{b}$. Lacking the birefringence in InGaP, phase matching is achieved by dispersion engineering of the InGaP nanophotonic waveguide. By designing the waveguide width, the effective index of the 1550 nm band TE$_{00}$ mode and the 775 nm band TM$_{00}$ mode can be equalized, as shown in Fig. \ref{fig:SHG}a using finite element simulation. 
For the phase-matched waveguide with length $L$, the normalized SHG efficiency, $\eta_{\mathrm{SHG}} \equiv\frac{P_{\mathrm{SHG}}}{P_p^2L^2}$, can be calculated using \cite{luo2018highly}
\begin{equation}
\begin{gathered}
    \eta_{\mathrm{SHG}} = \frac{\omega_a^2}{2n_a^2n_b\epsilon_0 c^3} \left( \frac{\int \,d\textbf{r} \chi_{xyz}^{(2)} \sum\limits_{i\neq j\neq k}E_{bi}^*E_{aj}E_{ak}}{\int \,d\textbf{r}\abs{\textbf{E}_a}^2\sqrt{\int \,d\textbf{r}\abs{\textbf{E}_b}^2}} \right) ^2,
\end{gathered}
\label{eq:PM}
\end{equation}
where $n_{a(b)}$ is the effective mode index of the fundamental(second)-harmonic mode and the normalization integrals use electric field components perpendicular to the wavevector of the waveguide mode. The SHG efficiency is optimized when the waveguide is aligned along the $(110)$ direction of InGaP, leading to simulated $\eta_{\mathrm{SHG}}=130,000\%$/W/cm$^2$ for the 1550 nm pump wavelength (SI). 

Fig. \ref{fig:SHG}b displays SEM images of a fabricated meander waveguide with a length of 1.6 mm. The waveguide connects to two adiabatically tapered couplers at the end to interface with tapered optical fibers for light transmission. The adiabatic coupler efficiency is approximately 80\% and 30\% for 1550 nm TE and 775 nm TM polarized light, respectively \cite{zhao2022ingap}. The waveguide width is tapered down before entering the $180^\degree$ turn to avoid mode interference due to the bending of the multimode waveguide. The SHG nonlinear transfer function of the meander waveguide is given by $\left(\frac{\sin N(x+\phi)}{N\sin(x+\phi)}\frac{\sin x}{x}\right)^2$ (Methods), where $x=\frac{\Delta k L_0}{2}$, $\Delta k=2k_\omega-k_{2\omega}$, $L_0$ is the length of the waveguide in one row, $N$ is the number of rows, and $2\phi$ is the total phase mismatch between the two modes in a $180^\degree$ turn. For phase-matched fundamental- and second-harmonic modes in the straight waveguide, they become phase mismatched in the turn because of the change of the waveguide width and waveguide bending. A narrow waveguide section with a tunable length (Fig. \ref{fig:SHG}b inset) is introduced to compensate the phase mismatch due to the bending waveguide such that the total phase mismatch $2\phi$ through the $180^\degree$ turn, including the tapering section, is multiple $2\pi$. As a result, the transfer function can be recovered to the ideal sinc$^2$ function.   

A tunable continuous-wave telecom band laser is employed for the measurement of second-harmonic generation. The output light from the waveguide passes through a 1550 nm/775 nm wavelength division multiplexer (WDM) to filter the residual pump before the second-harmonic intensity is measured. Fig. \ref{fig:SHG}c shows the SHG intensity of a 1.6 mm long waveguide as the pump wavelength is swept, where a peak on-chip SH power $P_{\mathrm{SHG}}=80\ \mu$W is observed for the pump wavelength 1542.1 nm and an on-chip pump power $P_p=1.56$ mW. This corresponds to a normalized SHG efficiency of 128,000\%/W/cm$^2$. In comparison, normalized SHG efficiency of 2,500\%/W/cm$^2$ was achieved in lossy InGaP waveguides with phase-matched higher-order modes before \cite{poulvellarie2021efficient}. A quadratic relationship between $P_{\mathrm{SHG}}$ and $P_p$ is observed in the pump non-depleted region (Fig. \ref{fig:SHG}d). 
Additionally, we explored the tunability of the waveguide's phase-matching condition through temperature tuning of the device. In Fig. \ref{fig:SHG}e, we present the measured SHG spectrum at several temperatures up to 95$^\circ$C, constrained by the thermoelectric cooler element. A tuning range of 12.7 nm and a temperature-induced shift of 0.17 nm/C in the phase-matching wavelength are measured.

The SHG efficiency realized in the InGaP nonlinear nanophotonic waveguide represents a substantial advance, in particular, in the crucial telecommunication band. Fig. \ref{fig:SHG}f displays the normalized SHG efficiency and 1550 nm band loss of the best telecommunication C band waveguide SHG, to our knowledge, of several $\chi^{(2)}$ nonlinear photonics platforms (a more comprehensive list is provided in SI).  The InGaP nanophotonic waveguide surpasses thin-film PPLN waveguides by nearly two orders of magnitude in terms of normalized nonlinear conversion efficiency \cite{wang2018ultrahigh,chen2024adapted}, while maintaining a low 1550 nm wavelength loss of 0.8$\pm$0.4 dB/cm, which is consistent with the measured $Q_i$ of microring resonators. The nonlinear efficiency of $P_{\mathrm{SHG}}/P_p^2=3280$\%/W achieved in the 1.6 mm long InGaP waveguide is comparable to the centimeter-long PPLN waveguide made with the adapted poling technique recently \cite{chen2024adapted}. For longer InGaP waveguides, we find the nonlinear efficiency deviates from the $L^2$ scaling because of the thickness nonuniformity of the thin film (SI). This can be mitigated using the adapted phase-matching technique \cite{chen2024adapted}, by varying the waveguide width according to the pre-calibrated InGaP film thickness to keep the phase-matching condition along the entire waveguide.

\begin{figure*}[!htb]
\begin{center}
\includegraphics[width=1.8\columnwidth]{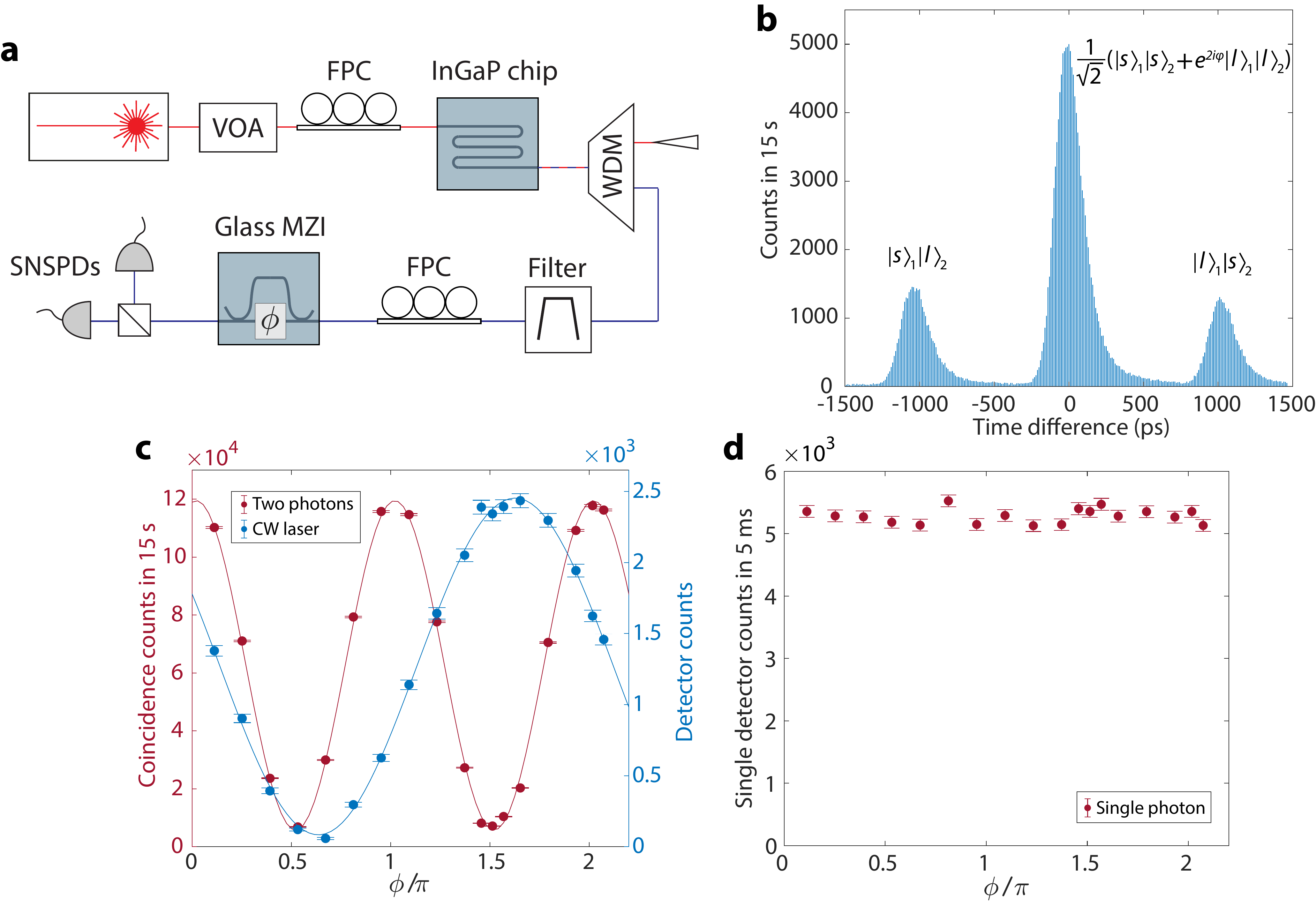}
\caption{\textbf{a}. Schematic of the measurement of the time-energy entanglement of the SPDC photons. VOA: variable optical attenuator. FPC: fiber polarization controller. \textbf{b}. Two-photon time difference histogram. The binwidth is 10 ps. The middle peak corresponds to the time-energy entangled state. \textbf{c}. Coincidence counts in 800 ps binwidth integrated in 15 s of the SPDC photon pair for various interferometer phases (red) and the interference fringe of a CW laser (blue). Solid lines are sinusoidal function fitting. Error bars represent the shot noise. \textbf{d}. Counts in 5 ms of SPDC photons in one SNSPD. }
\label{fig:entanglement}
\end{center}
\end{figure*}

Utilizing the InGaP nonlinear nanophotonic waveguide, we demonstrate an ultra-bright, broadband time-energy entangled photon source via spontaneous parametric down-conversion (SPDC). For a phase-matched nonlinear waveguide, the internal efficiency of the pair generation via SPDC can be related to the SHG efficiency by \cite{kumar2018parametric} 
\bqa\label{spdcpower}
P_{\mathrm{SPDC}}/P_p &\approx& \frac{\hbar\omega_pL^{3/2}}{3\sqrt{2\pi\left|\mathrm{GVD}\left(\frac{\omega_p}{2}\right)\right|}}\eta_{\mathrm{SHG}}\nonumber\\&=&\frac{\hbar\omega_pL^{2}\Delta f_{\mathrm{FWHM}}}{3\sqrt{2\pi}\alpha}\eta_{\mathrm{SHG}},
\eqa
where $\mathrm{GVD}\left(\frac{\omega_p}{2}\right)$ is the group velocity dispersion at $\frac{\omega_p}{2}$, $\alpha=\frac{1}{\pi}\sqrt{2\mathrm{sinc}^{-1}\frac{1}{\sqrt{2}}}$, and $\Delta f_{\mathrm{FWHM}}$ is the bandwidth of the SPDC photons given by 
\begin{equation}\label{bw}
\Delta f_{\mathrm{FWHM}} =\frac{\alpha}{\sqrt{\left|\mathrm{GVD}\left(\frac{\omega_p}{2}\right)\right|L}}.
\end{equation} 

We pumped the phase-matched waveguide with a tunable 780 nm band continuous-wave laser to generate telecommunication band SPDC photons. After filtering the residual pump, the SPDC photons were measured using either regular photodetectors or superconducting-nanowire single-photon detectors (SNSPDs). Fig. \ref{fig:SPDC}a shows the total photon pair generation rate efficiency via a 1.6 mm long waveguide measured with $P_p=135\ \mu$W. At the phase-matching wavelength of $772.12$ nm, a peak pair generation rate of 97 GHz/mW, corresponding to an internal efficiency of $2.5\times 10^{-5}$, is observed. 
To measure the bandwidth of the SPDC photons, we used a dense wavelength division-multiplexer (DWDM) with 40 channels and a 120 GHz channel bandwidth. The measured SPDC photon rate through each channel for $P_p=8.7\ \mu$W is displayed in Fig. \ref{fig:SPDC}b (the fiber coupling is not optimized for this measurement). The data is fitted using a $\mathrm{sinc}^2$ function and the SPDC photon bandwidth is inferred to be 14.4 THz (115 nm). The measured bandwidth agrees with the theoretical calculation using Eq. \ref{bw} (SI).
This leads to a per-bandwidth pair generation rate of 6.7 GHz/mW/THz (840 MHz/mW/nm). In Table \ref{tab:SPDC}, we compare the telecommunication C band SPDC photon pair generation in the InGaP waveguide and thin-film (TF) PPLN waveguides of several recent works. The InGaP waveguide SPDC source shows a rate efficiency at least an order of magnitude higher while retaining a large bandwidth. We also noticed a recent work of high-efficiency telecommunication L band SPDC photon generation in AlGaAs waveguides \cite{placke2024telecom}. A more comprehensive summary of broadband photon pair sources can be found, for example, in Ref. \cite{javid2021ultrabroadband}.

\renewcommand{\arraystretch}{1.2}
\begin{table}[htbp]
	\centering
    \caption{\textbf{Telecommunication C band SPDC photon pair generation in TF PPLN and InGaP waveguides}}
    \vspace{0pt}
    \resizebox{0.5\textwidth}{!}{
	\begin{tabular}{|c|c|c|c|}
		\hline
		Material & Rate (GHz/mW/THz) & Bandwidth (THz) & Waveguide length (mm)   \\
		\hline
		TF PPLN \cite{jin2014chip}   &   0.12 &  -- & 10 \\
		\hline
		TF PPLN \cite{zhao2020high}   &   0.46 &  -- & 5 \\
		\hline
		TF PPLN \cite{javid2021ultrabroadband}   & 0.13   &  100 & 5 \\
		\hline
		InGaP (this work)  &  6.7  & 14.4 & 1.6 \\
		\hline
	\end{tabular}}
	\label{tab:SPDC}
\end{table}
\vspace{6pt}

The second-order cross-correlation, $g_{\mathrm{si}}^{(2)}(\tau)$,  between the signal and idler photons via two DWDM channels was measured using a pair of SNSPDs. Fig. \ref{fig:SPDC}c displays the measured $g_{\mathrm{si}}^{(2)}(\tau)$ for pair generation rate of 18 MHz in one DWDM channel. In the low gain regime, the zero-delay cross-correlation between the signal and idler is shown to be \cite{clausen2014source}
\be\label{g2}
g_{\mathrm{si}}^{(2)}(0)=1+\frac{4B}{R}\frac{\Gamma_s\Gamma_i}{(\Gamma_s+\Gamma_i)^2},
\ee
where $R$ and $B$ are the total pair generation rate and bandwidth of the SPDC photons, respectively, and $\Gamma_{s,i}$ is the filter bandwidth of the signal and idler photons. For $\Gamma_{s}=\Gamma_{i}$, Eq. \ref{g2} indicates the inherent coincidence-to-accidental counts ratio (CAR) of the SPDC photon source is given by $\mathrm{CAR}=g_{\mathrm{si}}^{(2)}(0)-1=B/R$, i.e., the inverse of the photon pair rate per bandwidth. According to Eq. \ref{spdcpower} ($B\equiv \Delta f_{\mathrm{FWHM}}$ and $R\equiv P_{\mathrm{SPDC}}/\hbar\omega_p$), $\mathrm{CAR}\propto 1/(\eta_{\mathrm{SHG}}L^2P_p)$, which means for more efficient waveguides, same CAR can be achieved with less pump power. Due to the detector jitter ($\sim$ 100 ps), which is much larger than the coherence time of the SPDC photons filtered by a DWDM channel (120 GHz), the inherent $g_{\mathrm{si}}^{(2)}(0)$ cannot be resolved and the measured CAR will be lower than the inherent value \cite{clausen2014source}.
The measured CAR and pair generation rate in one DWDM channel for various pump power is shown in Fig. \ref{fig:SPDC}d (Methods). Nevertheless, CAR $>10^4$ is observed for relatively low photon pair generation rate. 

To demonstrate the time-energy entanglement of the SPDC photons, we measured the two-photon interference using an unbalanced Mach-Zehnder interferometer (MZI) \cite{brendel1991time}, as illustrated in Fig. \ref{fig:entanglement}a  (Methods). The unbalanced MZI is made from glass-based photonic integrated circuits with a path delay of $\tau_d=1$ ns. The SPDC photons used in this measurement have a bandwidth of about 20 nm as they are filtered by a CWDM, leading to a single-photon coherence time much shorter than $\tau_d$. The coherence time of the correlated signal-idler pair is determined by the continuous-wave pump laser, which is much longer than $\tau_d$. As a result, the signal-idler photon pair can interfere through the unbalanced MZI while neither can the signal or idler single photon. The signal-idler pair can travel through either the short ($s$) or long ($l$) path of the unbalanced MZI together, forming a time-energy entangled state $\ket\psi=\frac{1}{\sqrt{2}}(\ket s_1\ket s_2+e^{2i\phi}\ket l_1\ket l_2)$, where $\phi$ is the phase difference between the two paths for light with a frequency that is half of the SPDC pump frequency. This entangled state can be post-selected, distinguishing itself from the other two states out of the interferometer, $\ket s_1\ket l_2$ and $\ket l_1\ket s_2$, using time-resolved coincidence detection (Methods). 

Fig. \ref{fig:entanglement}b shows a measured time difference histogram of the SPDC photons. The coincidence counts around the zero delay, corresponding to the entangled state, depend on the interferometer phase as $\propto\frac{1}{2}(1+\cos2\phi)$. Fig. \ref{fig:entanglement}c shows the measured two-photon interference fringe by varying the temperature of the glass interferometer which changes $\phi$.  The two-photon fringe has a period of $\pi$ while the fringe of a continuous-wave laser with a frequency about half of the pump frequency has a period of $2\pi$. The counts of SPDC photons in one detector does not exhibit an interference fringe as expected (Fig. \ref{fig:entanglement}d). The measured two-photon interference visibility is 90.8\%, which is limited by the imperfect glass photonic circuit MZI with the beam splitter ratio deviating from 50/50. The two-photon interference visibility is $\geq98.6\%$ after correction for the interferometer imperfection (SI). The two-photon interference visibility exceeds the Clauser-Horne limit of $\frac{1}{\sqrt{2}}\approx70.7\%$, which proves the photon pair entanglement \cite{clauser1974experimental}. 

\noindent\textbf{Discussion}\\
In summary, we have demonstrated a broadband second-order nonlinear photonics platform based on thin-film InGaP. With the optimized optical loss and phase-matching condition, the InGaP nanophotonic waveguide enables second-order nonlinear optical processes, including SHG and SPDC, with normalized efficiencies one to two orders of magnitude higher than the state of the art in the telecommunication C band. The nonlinear efficiency of the InGaP waveguides can be further enhanced using the adapted fabrication technique \cite{chen2024adapted} to counter the thickness nonuniformity of thin films. The ultra-bright, broadband entangled photon pair source, covering the telecommunication C band, will be useful for high-rate wavelength-multiplexed entanglement distribution over long distances \cite{wengerowsky2018entanglement} and ultrafast spectroscopy using entangled photons \cite{zhang2022entangled}. Beyond that, the demonstrated InGaP nonlinear photonics platform is expected to enable unprecedented performances in applications ranging from squeezed light generation \cite{nehra2022few,stokowski2023integrated} and optical parametric amplification \cite{riemensberger2022photonic} to few-photon quantum nonlinear optics \cite{yanagimoto2022temporal}, among others. Based on  III-V semiconductors, the InGaP platform also enables monolithic integration of pump lasers \cite{majid2015first} to realize electrically-injected nonlinear photonic sources.

\noindent\textbf{Materials and methods}\\
\textbf{Device fabrication}\\ The device fabrication follows Ref. \cite{zhao2022ingap}. The device pattern is defined using electron beam lithography and transferred to InGaP layer via inductively coupled plasma reactive-ion etch (ICP-RIE) using a mixture of Cl$_2$/CH$_4$/Ar gas. Then a layer of aluminum oxide is deposited via atomic layer deposition. The InGaP device is released from the GaAs substrate using citric acid-based selective etching.\\
\textbf{Measurement} \\ For the measurement of SPDC photons, we used a 780 nm band continuous-wave tunable diode laser as the pump. The light is polarization aligned by a fiber polarization controller, coupled into the device via a tapered fiber to generate the SPDC photons, and coupled back into the optical fiber using another tapered fiber. The residual pump is filtered by a 1550 nm/780 nm WDM. To measure the cross-correlation and CAR of the SPDC photons, the signal and idler photons are separated by a DWDM and detected using SNSPDs (Quantum Opus) and time-correlated single-photon counting module (Swabian). The coincidence and accidental counts are integrated in a time binwidth of 10 ps.  To measure the time-energy entanglement of the SPDC photons, both signal and idler photons are filtered by the same CWDM channel and then pass through a glass waveguide unbalanced MZI (Teem Photonics). The two-photon coincidences are detected by two SNSPDs after the 50/50 beam splitter. \\
\textbf{Nonlinear transfer function} \\
The transfer function of SHG is calculated as 
\be
\mathcal{F}(L)=\left|\frac{1}{L}\int_0^L e^{i\int_0^z\Delta \phi dz'}dz\right|^2,
\ee
where $\Delta \phi$ is the phase mismatch between the fundamental- and second-harmonic modes. For a meander waveguide, 
\bqa
\mathcal{F}(L)&=&\left|\frac{1}{L}\sum_n\int_{nL_0}^{(n+1)L_0} e^{i(\Delta k z+2n\phi)}dz\right|^2\\\nonumber
&=&\left|\frac{e^{i\Delta kL_0}-1}{i\Delta kL}\frac{e^{iN(\Delta kL_0+2\phi)}-1}{e^{i(\Delta kL_0+2\phi)}-1}\right|^2\\\nonumber
&=&\left(\frac{\sin(\Delta kL_0/2)}{\Delta kL/2}\frac{\sin\big(N(\Delta kL_0/2+\phi)\big)}{\sin(\Delta kL_0/2+\phi)}\right)^2,
\eqa
where $\Delta k=2k_\omega-k_{2\omega}$, $L_0$ is the length of the waveguide in one row, $N$ is the number of rows, and $2\phi$ is the total phase mismatch between the two modes for a $180^\degree$ turn.

\vspace{2mm}
\noindent\textbf{Acknowledgements}\\ 
This work was supported by US National Science Foundation (Grant No. ECCS-2223192), NSF Quantum Leap Challenge Institute QLCI-HQAN (Grant No. 2016136), and U.S. Department of Energy Office of Science National Quantum Information Science Research Centers. 

\noindent\textbf{Author contributions}\\ 
Y.Z., J.A., and A.K.M.N.H. performed the simulation. J.A. and Y.M. fabricated the device. J.A., Y.Z., Y.M., and A.K.M.N.H. measured the device and analyzed the data. J.A., Y.Z., and K.F. wrote the manuscript.

\noindent\textbf{Conflict of interest}\\ 
The authors declare no competing interests.

 \vspace{10mm}



%

\end{document}


\title{Supplementary Information for: ``InGaP $\chi^{(2)}$ integrated photonics platform for broadband, ultra-efficient nonlinear conversion and entangled photon generation'' }

\author{Joshua Akin} 
\thanks{These authors contributed equally to this work.}
\affiliation{Holonyak Micro and Nanotechnology Laboratory and Department of Electrical and Computer Engineering, University of Illinois at Urbana-Champaign, Urbana, IL 61801 USA}
\affiliation{Illinois Quantum Information Science and Technology Center, University of Illinois at Urbana-Champaign, Urbana, IL 61801 USA}
\author{Yunlei Zhao} 
\thanks{These authors contributed equally to this work.}
\affiliation{Holonyak Micro and Nanotechnology Laboratory and Department of Electrical and Computer Engineering, University of Illinois at Urbana-Champaign, Urbana, IL 61801 USA}
\affiliation{Illinois Quantum Information Science and Technology Center, University of Illinois at Urbana-Champaign, Urbana, IL 61801 USA}
\author{Yuvraj Misra} 
\affiliation{Holonyak Micro and Nanotechnology Laboratory and Department of Electrical and Computer Engineering, University of Illinois at Urbana-Champaign, Urbana, IL 61801 USA}
\affiliation{Illinois Quantum Information Science and Technology Center, University of Illinois at Urbana-Champaign, Urbana, IL 61801 USA}
\author{A. K. M. Naziul Haque} 
\affiliation{Holonyak Micro and Nanotechnology Laboratory and Department of Electrical and Computer Engineering, University of Illinois at Urbana-Champaign, Urbana, IL 61801 USA}
\affiliation{Illinois Quantum Information Science and Technology Center, University of Illinois at Urbana-Champaign, Urbana, IL 61801 USA}
\author{Kejie Fang} 
\email{kfang3@illinois.edu}
\affiliation{Holonyak Micro and Nanotechnology Laboratory and Department of Electrical and Computer Engineering, University of Illinois at Urbana-Champaign, Urbana, IL 61801 USA}
\affiliation{Illinois Quantum Information Science and Technology Center, University of Illinois at Urbana-Champaign, Urbana, IL 61801 USA}

\maketitle

\section{Microring quality factor}

\begin{figure*}[!htb]
\begin{center}
\includegraphics[width=0.4\columnwidth]{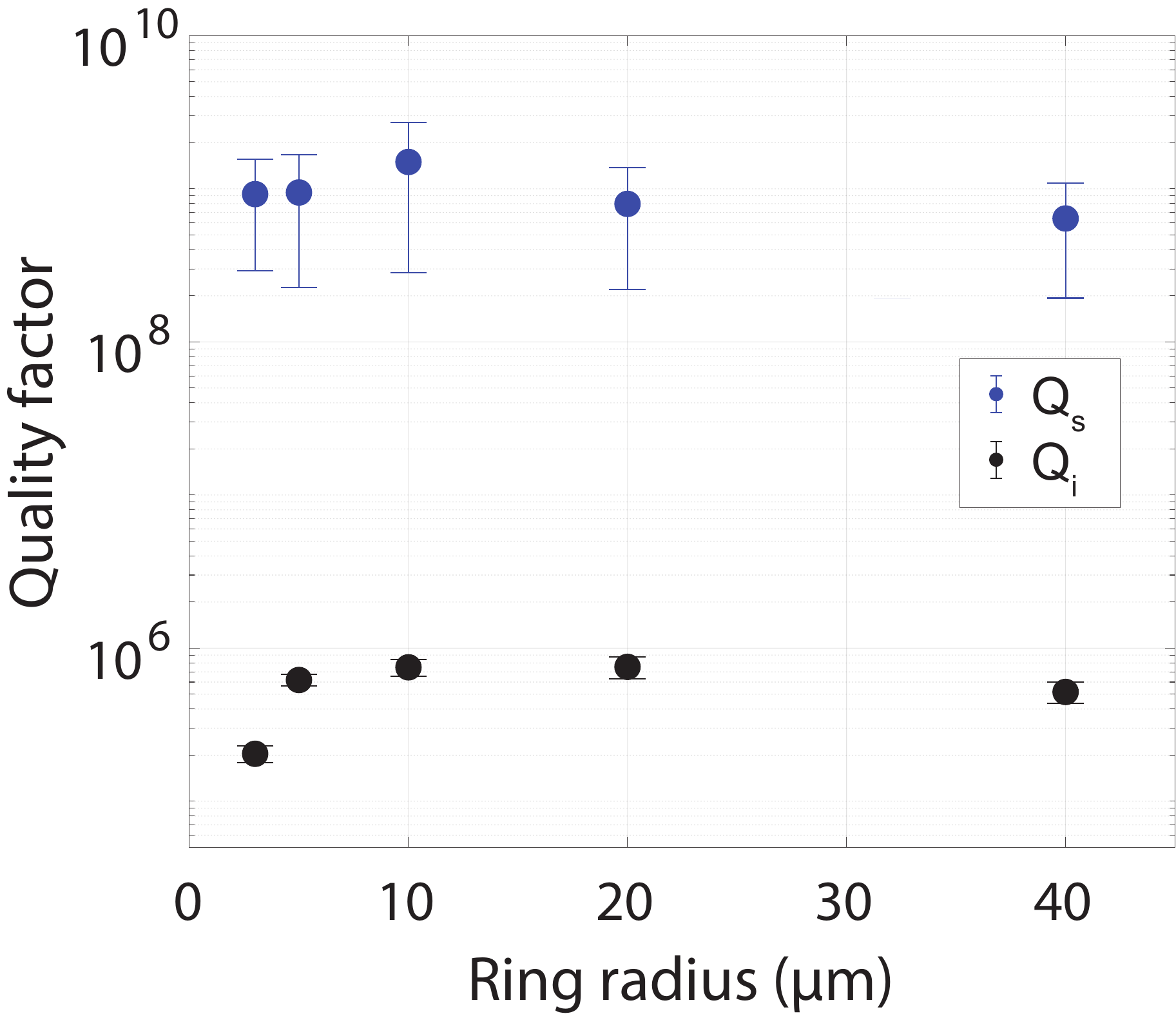}
\caption{ Quality factor of the 1550 nm band TE$_{00}$ resonances of microring resonators of different radius. }
\label{figs1:ringq}
\end{center}
\end{figure*}

Microring resonators of different radius are made and the quality factor of the 1550 nm band TE$_{00}$ resonances are measured. The result is displayed in Fig. \ref{figs1:ringq}. The intrinsic quality factor of the microring resonator is found to be bounded around $8\times 10^5$, which is limited by the absorption loss. The intrinsic quality factor can be decomposed as $1/Q_i=1/Q_r+1/Q_s+1/Q_a$, where $1/Q_r$ is the radiation loss due to the curvature, $1/Q_s$ is the surface roughness caused scattering loss, $1/Q_a$ is the absorption loss. From numerical simulation, the radiation loss of microrings with radius larger than 5 $\mu$m is negligible. Below we calculate the scattering loss using the model of Ref. \cite{borselli2005beyond} for the microring resonators. 

The splitting of the microring resonances due to surface roughness induced backscattering is given by 
\be\label{sieq1}
\frac{\Delta\lambda }{\lambda}=2\sqrt{2} \pi^{3/4} \xi\frac{V_s}{V_r},
\ee 
where $V_r$ is the physical volume of the microring, $V_s$ is the effective volume of a typical scatterer, and $\xi$ is the relative dielectric contrast
constant defined as 
\be
\xi=\frac{\overline{n}^2(n_r^2-n_0^2)}{n_r^2(\overline{n}^2-n_0^2)}.
\ee
$n_r$, $n_0$, and $\bar n$ are the indices of refraction for the InGaP microring, surrounding medium, and 2D effective slab, respectively. 

The surface roughness induced scattering loss $Q_s$ is given by
\be\label{sieq2}
Q_{s}=\frac{3 \lambda^3}{16 \pi^{7/2} n_0 (n_r^2-n_0^2) \xi}\frac{V_r}{V_s^2}.
\ee
From Eqs. \ref{sieq1} and \ref{sieq2}, we can calculate $Q_s$ using the measured resonance splitting $\Delta \lambda$. $Q_s$ is found to be on the order of $10^9$, as shown in Fig. \ref{figs1:ringq}, which is much larger than the measured $Q_i$. As a result, we conclude $Q_i$ is limited by the material absorption loss. 

\section{Second Harmonic Generation Efficiency}
For an InGaP waveguide with phase-matched fundamental-harmonic (FH) mode $a$ and second-harmonic (SH) mode $b$, the SHG efficiency can be calculated using \cite{luo2018highly}:
\begin{equation}
	\begin{gathered}
		\eta_{\mathrm{SHG}} = \frac{\omega_a^2}{2n_a^2n_b\epsilon_0 c^3} \left( \frac{\int \,d\textbf{r} \chi_{xyz}^{(2)} \sum\limits_{i\neq j\neq k}E_{bi}^*E_{aj}E_{ak}}{\int \,d\textbf{r}\abs{\textbf{E}_a}^2\sqrt{\int \,d\textbf{r}\abs{\textbf{E}_b}^2}} \right) ^2,
	\end{gathered}
	\label{eq:PM}
\end{equation}
with the normalization integrals only considering the transverse field components. Because the second-order susceptibility of InGaP only has the term $\chi_{xyz}^{(2)}$, the FH and SH modes are chosen to be TE$_{00}$ and TM$_{00}$, respectively. Consider the orientation of the waveguide has an angle of $\theta$ with respect to the $(100)$ direction of InGaP. We denote the electric field in the waveguide frame as $(E_{x'}, E_{y'}, E_{z'})$, where the $x'-z'$ plane is the waveguide cross-section plane and $y'$ direction is along the waveguide. The electric field in the crystal frame is denoted as $(E_{x}, E_{y}, E_{z})$. As a result, $E_x=E_{x'}\cos\theta+E_{y'}\sin\theta$, $E_y=E_{x'}\sin\theta-E_{y'}\cos\theta$, and $E_z=E_{z'}$. Because the FH and SH modes are TE$_{00}$ and TM$_{00}$, respectively, the dominant contribution to the numerator integral of Eq. \ref{eq:PM} is given by
\bqa
&&\int \,d\textbf{r} \chi_{xyz}^{(2)} \sum\limits_{i\neq j\neq k}E_{bi}^*E_{aj}E_{ak}\\\nonumber
&=&2\int \,d\textbf{r} \chi_{xyz}^{(2)} E_{bz}^*E_{ax}E_{ay}\\\nonumber
&=&2\int \,d\textbf{r} \chi_{xyz}^{(2)} E_{bz'}^*(E_{ax'}\cos\theta+E_{ay'}\sin\theta)(E_{ax'}\sin\theta-E_{ay'}\cos\theta)\\\nonumber
&=&\int \,d\textbf{r} \chi_{xyz}^{(2)} E_{bz'}^*(E_{ax'}^2-E_{ay'}^2)\sin2\theta-2\int \,d\textbf{r} \chi_{xyz}^{(2)} E_{bz'}^*E_{ax'}E_{ay'}\cos2\theta\\\nonumber
&=&\int \,d\textbf{r} \chi_{xyz}^{(2)} E_{bz'}^*(E_{ax'}^2-E_{ay'}^2)\sin2\theta.\\\nonumber
\eqa
The second term in the second-last line is zero because $E_{ay'}$ is odd while $E_{ax'}$ and $E_{bz'}$ are even.
Other terms contributing to the nonlinear interaction include
\bqa
&&2\int \,d\textbf{r} \chi_{xyz}^{(2)} E_{az}(E_{bx}^*E_{ay}+E_{by}^*E_{ax})\\\nonumber
&=&2\int \,d\textbf{r} \chi_{xyz}^{(2)} E_{az'}(E_{bx'}^*\cos\theta+E_{by'}^*\sin\theta)(E_{ax'}\sin\theta-E_{ay'}\cos\theta)\\\nonumber
&&+2\int \,d\textbf{r} \chi_{xyz}^{(2)} E_{az'}(E_{bx'}^*\sin\theta-E_{by'}^*\cos\theta)(E_{ax'}\cos\theta+E_{ay'}\sin\theta)\\\nonumber
&=&2\int \,d\textbf{r} \chi_{xyz}^{(2)} E_{az'}(E_{bx'}^*E_{ax'}\sin2\theta-E_{bx'}^*E_{ay'}\cos2\theta-E_{by'}^*E_{ax'}\cos2\theta-E_{by'}^*E_{ay'}\sin2\theta)\\\nonumber
&=&0,
\eqa
which vanish because of the symmetry of the modes.

 As a result, the mode overlap integral is maximized when $\theta=\frac{\pi}{4}$. The optimized SHG efficiency thus is given by
\begin{equation}
	\begin{gathered}
		\eta_{\mathrm{SHG}} = \frac{\omega_a^2}{2n_a^2n_b\epsilon_0 c^3} \left( \frac{\int \,d\textbf{r} \chi_{xyz}^{(2)} E_{bz'}^*(E_{ax'}^2-E_{ay'}^2)}{\int \,d\textbf{r}\abs{\textbf{E}_a}^2\sqrt{\int \,d\textbf{r}\abs{\textbf{E}_b}^2}} \right) ^2.
	\end{gathered}
\end{equation}
The $E$ field of the FH and SH modes are obtained by finite element simulation and the optimized SHG efficiency for 110 nm thick InGaP phase-matched waveguide and 1550 nm FH wavelength is found to be $\eta_{\mathrm{SHG}}=130,000\%$/W/cm$^2$, which agrees with the measured value. 

\section{Nonlinear efficiency of InGaP waveguides}

\begin{figure*}[!htb]
\begin{center}
\includegraphics[width=0.8\columnwidth]{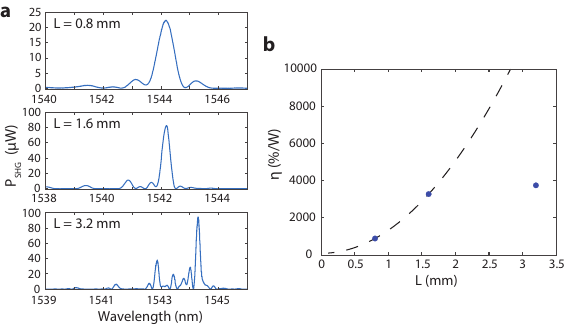}
\caption{ \textbf{a}-\textbf{c}. Measured SHG spectrum of three waveguides with length 0.8, 1.6, and 3.2 mm, respectively. \textbf{d}. Peak nonlinear efficiency v. waveguide length. Dashed line indicates the $L^2$ scaling. }
\label{figs2:nonuniformity}
\end{center}
\end{figure*}

The nonlinear efficiency of long InGaP waveguides is limited by the thickness nonuniformity of the thin film. Fig. \ref{figs2:nonuniformity}a-c display the SHG spectrum of three waveguides with length 0.8, 1.6, and 3.2 mm, respectively, measured with $P_p=1.56$ mW. It is observed that the spectrum of the 3.2 mm long waveguide distorts more due to the thickness nonuniformity as well as the phase mismatch at the $180^\degree$ waveguide turn. Fig. \ref{figs2:nonuniformity}d shows the peak nonlinear efficiency of the three waveguides. The nonlinear efficiency of the 3.2 mm long waveguide deviates from the $L^2$ scaling. 

\section{Bandwidth of waveguide SPDC photons}
The SPDC signal power in a bandwidth of $d\omega_s$ is given by \cite{kumar2018parametric}

\begin{equation}\label{signalPower}
	dP_s = \frac{\hbar d_{\mathrm{eff}}^2P_p\omega_s^2\omega_iL^2}{\pi\epsilon_0c^3n_sn_in_pA_{\mathrm{eff}}}\mathrm{sinc}^2\left( \frac{\Delta kL}{2}\right)d\omega_s,
\end{equation}
where the $d_{\mathrm{eff}}$ is the nonlinear coefficient, $L$ is the waveguide length, $P_p$ is the pump power, $A_{\mathrm{eff}}$ is the effective interaction area, and $\Delta k$ is the phase mismatch between the pump, signal and idler defined as
	\begin{equation}\label{phaseMismatch}
		\Delta k = \frac{\left( n_p\omega_p - n_s\omega_s - n_i\omega_i\right)}{c}.
	\end{equation}
Eq. \ref{signalPower} shows the ideal sinc$^2$ function of the SPDC  spectrum and the bandwidth of the SPDC photons can be calculated from Eq. \ref{phaseMismatch}.
	
The InGaP waveguide is designed to be perfectly phase matched for the degenerate signal and idler. Thus the nondegenerate signal and idler frequencies are expressed most conveniently as
	\begin{equation}\label{siFrequencies}
		\omega_{s,i} = \frac{\omega_p}{2} \pm \Delta \omega.
	\end{equation}
	Using Eq. \ref{phaseMismatch}, the phase mismatch can be expressed in terms of the group velocity dispersion:
	\begin{equation}\label{GVD}
			\Delta k = \mathrm{GVD}\left(\frac{\omega_p}{2}\right)\Delta \omega^2,
				\end{equation}
where the group velocity dispersion is defined as
			\begin{equation}
			\mathrm{GVD}(\omega_0) \equiv \frac{2}{c}\left(\frac{\partial n}{\partial\omega}\right)_{\omega = \omega_0} +\frac{\omega_0}{c}\left(\frac{\partial^2 n}{\partial\omega^2}\right)_{\omega = \omega_0}.
	\end{equation}
	$n(\omega)$ is the effective mode index of the waveguide mode corresponding to the signal and idler.
Using Eqs. \ref{signalPower} and \ref{GVD}, the full-width-half-maximum (FWHM) bandwidth of the SPDC spectrum is then given by
	\begin{equation}\label{spdcBandwidth}
		\Delta f_{\mathrm{FWHM}} = \frac{\alpha}{\sqrt{\vert \mathrm{GVD}\left(\frac{\omega_p}{2}\right)\vert L}},
	\end{equation}
	where $\alpha = \frac{1}{\pi}\sqrt{2\mathrm{sinc}^{-1}\frac{1}{\sqrt{2}}}$.
	
	For the waveguide used for the SPDC process, $L=1.6$ mm and the pump frequency is $\omega_p/2\pi=388.55$ THz. We simulated the dispersion of the effective mode index of the TE$_{00}$ waveguide mode around $\omega_p/2$ and found $\vert\mathrm{GVD}\left(\frac{\omega_p}{2}\right)\vert = 0.373\ (\mathrm{ps})^2/\mathrm{m}$. This yields a SPDC bandwidth of 12.1 THz, which is close to the measured value.

\section{Two-photon interference visibility}
	Consider an unbalanced Mach-Zehnder interferometer (MZI) consisting of two beam splitters, with transmission and reflection coefficients $T_{1(2)}$ and $R_{1(2)}$, which satisfy $T_k^2  + R_k^2 = 1$. For a single-photon input state, the output state of the MZI is given by
	\begin{equation}
		\ket 1 \rightarrow T_1T_2\ket s+ R_1 R_2 e^{i \phi}\ket l,
	\end{equation}
where $\ket s$ and $\ket l$ represent the state in the short and long arm of the unbalanced MZI, respectively, and $\phi$ is the phase difference of the two arms. The visibility of the single-photon interference fringe thus is
	\begin{equation}\label{singlephoton}
		V_1 = \frac{2T_1 T_2 R_1 R_2}{T_1^2 T_2^2 + R_1^2 R_2^2}.
	\end{equation}
The single-photon interference visibility can be characterized using a CW laser.
	
	For a correlated two-photon input state $\ket 1_1\ket 1_2$ generated via SPDC, without parasitic noises, the output state thus is given by
	\begin{equation}
		\ket 1_1\ket 1_2 \rightarrow T_1^2 T_2^2 \ket s_1 \ket s_2 + T_1 T_2 R_1 R_2 e^{i \phi} \ket s_1 \ket l_2 + T_1 T_2 R_1 R_2 e^{i \phi} \ket l_1 \ket s_2 + R_1^2 R_2^2 e^{2i \phi} \ket l_1 \ket l_2,
	\end{equation}
where $\phi$ is the interferometer phase difference corresponding to half of the SPDC pump frequency.
	In this output state, only the first and last terms contribute to the two-photon interference, since they cannot be distinguished in the two-photon coincidence measurement. The visibility of the two-photon interference is given by
	\begin{equation}\label{twophoton}
		V_2=\frac{2T_1^2 T_2^2 R_1^2 R_2^2}{T_1^4 T_2^4 + R_1^4 R_2^4}.
	\end{equation}
Using Eqs. \ref{singlephoton} and \ref{twophoton}, the two interference visibilities are related by
\be\label{v1v2}
V_2=\frac{2}{\frac{4}{V_1^2}-2}.
\ee
	
We model the measured two-photon interference visibility $V_{2,m}$ deviating from the ideal case (Eq. \ref{v1v2}) using some uncorrelated counts $x$. We stress this is an effective model and it could represent other physcal imperfection in the setup. The minimum and maximum two-photon coincidence counts are given by
$C_{\mathrm{min}}=(T_1^2 T_2^2-R_1^2 R_2^2)^2+x$ and $C_{\mathrm{max}}=(T_1^2 T_2^2+R_1^2 R_2^2)^2+x$. The measured two-photon interference visibility thus is
	\begin{equation}\label{measuredv2}
		V_{2,m}=\frac{2T_1^2 T_2^2 R_1^2 R_2^2}{T_1^4 T_2^4 + R_1^4 R_2^4+x}.
	\end{equation}
Using Eqs. \ref{singlephoton} and \ref{measuredv2}, we can find the bound of $x$:
\bqa\label{xmax}
x&=&\frac{2}{V_{2,m}}T_1^2 T_2^2 R_1^2 R_2^2-T_1^4 T_2^4 - R_1^4 R_2^4\\\nonumber
&=&T_1^2 T_2^2 R_1^2 R_2^2\left(\frac{2}{V_{2,m}}-\frac{4}{V_1^2}+2\right)\\\nonumber
&=&\mu^2\left(\frac{(1-R_1^2)R_1^2}{(\mu^2-1)R_1^2+1}\right)^2\left(\frac{2}{V_{2,m}}-\frac{4}{V_1^2}+2\right)\\\nonumber
&\leq&\frac{\mu^2}{(\mu+1)^4}\left(\frac{2}{V_{2,m}}-\frac{4}{V_1^2}+2\right)\\\nonumber
&=&x_{\mathrm{max}},
\eqa
where $\mu=(1+\sqrt{1-V_1^2})/V_1$. In our experiment, the measured CW interference visibility and two-photon interference visibility are $V_1=97.9\%$ and $V_{2,m}=90.8\%$, respectively, from which we find $x_{\mathrm{max}}=0.00179$.

We can then estimate the inherent two-photon interference visibility, $V_{2,i}$, if we were to use a perfect unbalanced MZI with $T_k^2=R_k^2=\frac{1}{2}$,
\be\label{v2i}
		V_{2,i}=\frac{\frac{1}{8}}{\frac{1}{8}+x}\geq\frac{\frac{1}{8}}{\frac{1}{8}+x_{\mathrm{max}}}=98.6\%.
\ee
If we were to use a practically better interferometer with $V_1=99.8\%$ (achieved with $T_k^2=0.484$ and $R_k^2=0.516$), then using Eq. \ref{measuredv2} and the estimated bound $x_{\textrm{max}}$, the measured two-photon interference visibility will be $V_{2,m}\geq97.8\%$.

\section{Waveguide SHG in thin-film $\chi^{(2)}$ photonics platforms}
	
In Table \ref{tab:cmpr}, we summarize results, to our knowledge, of thin-film waveguide SHG in the telecommunication and near-infrared wavelength bands from recent works. The top half of the table before the break is in the telecommunication C band. 

\begin{table*}[!htb]
	\centering
	\caption{\textbf{Thin-film waveguide SHG performances}}
	\small
	\begin{tabular}{|c|c|c|c|c|c|c|}
		\hline
		 Material & $\chi^{(2)}$ (pm/V) & poling & pump wavelength (nm) & $\eta_{\mathrm{SHG}}$ (W$^{-1}$cm$^{-2}$) & $\eta$ (W$^{-1}$)     & FH loss (dB/cm)\\
		\hline
            InGaP (this work) &$\chi^{(2)}_{xyz}$ = 220&no	&	1550 &	 128000$\%$& 3280$\%$ 	&	0.8
            \\
            \hline
            InGaP \cite{poulvellarie2021efficient} &$\chi^{(2)}_{xyz}$ = 220&no	&	1536 &	 2500$\%$& 12$\%$ 	&	 12
            \\
            \hline
            PGLN \cite{wang2017second}&	$\chi^{(2)}_{zzz}$ = 54 &	no&  1550&	60$\%$&	31$\%$&		3
		\\
		\hline
            LN \cite{luo2018highly} &	 $\chi^{(2)}_{zzx}$ = 8.6& no& 1540&	22$\%$&	5$\%$ 	&		0.5
            \\
		\hline
		PPLN \cite{wang2018ultrahigh}&	$\chi^{(2)}_{zzz}$ = 54 &	yes&  1550&	2600$\%$&	42$\%$  &		 $\sim$ 1
            \\
		\hline
            PPLN \cite{stokowski2023integrated} &	$\chi^{(2)}_{zzz}$ = 54&	yes&1544&	1000\%&	660\%&		0.7
		\\
		\hline
		PPLN \cite{chen2024adapted} &	$\chi^{(2)}_{zzz}$ = 54&	yes&1530&	2000\%&	9000\%&		0.8
		\\
		\hline
            Al$_{0.27}$Ga$_{0.73}$As \cite{may2019second}&	$\chi^{(2)}_{xyz} =$ 210 &	no &1560&	1202$\%$&	87$\%$ &	14
		\\
		\hline\hline
		          Al$_{0.19}$Ga$_{0.81}$As \cite{roland2020second}&	$\chi^{(2)}_{xyz}=$ 220&	no  & 1590 &	1600$\%$&	16$\%$ &	15	
		\\
		\hline
		 Al$_{0.2}$Ga$_{0.8}$As \cite{placke2024telecom}&	$\chi^{(2)}_{xyz} =$ 210 &	no &1580&	23000$\%$&	477$\%$ & {$\leq$ 2}
		\\
		\hline
            GaP \cite{pantzas2022continuous}&	$\chi^{(2)}_{yxx}$ = 70 &	yes & 1595&	200$\%$ &	14$\%$ &		8
            \\
		\hline
            SiC \cite{zheng2022efficient}& $\chi^{(2)}_{zzz}=$ 25 &no& 1584&	60$\%$&	5$\%$ &0.4 \cite{ou2023novel}
            \\
            \hline
            GaAs \cite{chang2018heterogeneously}&	$\chi^{(2)}_{xyz}=$ 240 &	no&  2025&	13000$\%$&	250$\%$  &		2
            \\
		\hline
            GaAs \cite{stanton2020efficient}&	$\chi^{(2)}_{xyz}$ = 240&	no & 1968 &	47600$\%$ &	4000$\%$ &		1.5
            \\
		\hline
	\end{tabular}%
	\label{tab:cmpr}
\end{table*}

\newpage

\vspace{2mm}

%